\documentstyle[epsf,preprint,aps]{revtex}
\begin{document}
\draft

\def\be{\begin{equation}}
\def\ee{\end{equation}}
\def\bea{\begin{eqnarray}}
\def\eea{\end{eqnarray}}
\def\bn{{\bf n}}
\def\bp{{\bf p}}
\def\bpi{\mbox{\boldmath $\pi$}}
\def\r{\rho}
\def\s{\sigma}
\def\t{\theta}
\def\nint{\int \!\! d\bn}
\def\ro{\frac{n-1}{2}}
\def\roo{(n-1)/2}
\def\rooo{\frac{n}{2} - 1}
\def\tint{\int_{0}^{\infty} \! d\t}
\def\intt{\int_{-\infty}^{\infty} \! dt}
\def\lint{\int_{0}^{\infty} \! d\l \mu(\l)}
\def\a{\frac{1}{2 K_{\roo}(J)}}
\def\as{a \, S_{n-1}}
\def\argc{J \cosh\t \cosh\t'}
\def\args{J \sinh\t \sinh\t'}
\def\l{\lambda}
\def\L{\Lambda}
\def\G{\Gamma}
\def\la{\langle}
\def\ra{\rangle}

\centerline{\bf Phase Transition in a Model with Non-Compact Symmetry 
on Bethe Lattice}
\centerline{\bf  and the
Replica Limit}
\vskip 1cm
\centerline{Ilya A. Gruzberg}
\centerline{\it Physics Department, Yale University, New Haven, Connecticut 
06511, U.S.A.}
\vskip .5cm
\centerline{Alexander D. Mirlin \footnote
{On leave from the Petersburg Nuclear Physics Institute, 188350
Gatchina, St.Petersburg, Russia}}
\begin{center}
{\it Insitut f\"{u}r Theorie der Kondensierten Materie,\\
Universit\"{a}t Karlsruhe, 76128 Karlsruhe, Germany}
\end{center}


\begin{abstract}

We solve $O(n,1)$ nonlinear vector model on Bethe lattice and show that it
exhibits a transition from ordered to disordered state for $0 \leq n < 1$.
If the replica limit $n\to 0$ is taken carefully, the model is shown to
reduce to the corresponding supersymmetric model.  The latter was
introduced by Zirnbauer as a toy model for the Anderson localization
transition.  We argue thus that the non-compact replica models describe
correctly the Anderson transition features.  This should be contrasted to
their failure in the case of the level correlation problem.

\end{abstract}


\pagebreak

\section{Introduction}
\label{sec:intro}

In the study of systems with quenched randomness historically the first way
of treating the disorder averages was the so called replica method of
Edwards and Anderson \cite{EA75}.  In this method one introduces $n$
``replicas'' of the original system and calculates annealed averages for
this replicated system.  Then the use of identity
\[
\overline {\log Z} = \lim_{n \rightarrow 0} \frac{\overline {Z^{n}} -
1}{n}
\]
allows to recover the properties of the original system.

For the problem of Anderson localization replica fields describing
electrons may be chosen to be either fermionic (Grassmann) or bosonic. In
either case one can construct effective field theoretic description in
terms of non-linear $\s$-model for matrix field which belongs to some
compact manifold for fermionic replicas and some non-compact manifold for
bosonic ones (see, e.g.  \cite{MS81}).

Another way to treat the disorder is the supersymmetry method of Efetov
\cite{E83}.  In this method one introduces both fermionic and bosonic
degrees of freedom and the resulting $\s$-model field is of the supermatrix
structure.

It was realized some time ago that the replica method suffers from serious
drawbacks.  Verbaarschot and Zirnbauer showed explicitly \cite{VZ85} that
the replica method fails to give correct non-perturbative result for a
problem of energy level correlation which is equivalent to zero-dimensional
$\s$-model, while the supersymmetry method works nicely.  Since then the
prevailing opinion in the literature on the Anderson localization problem
was that the replica method may be at best considered as perturbative tool
not being able to describe properties of disordered (localized) phase.

The goal of our paper is to reconsider the correspondence between
supersymmetric models and non-compact models in the replica limit. For
this purpose we analyze the solution of two ``toy'' models on Bethe
lattice.  The first one is one of the simplest models with non-compact
symmetry, namely $O(n,1)/O(n)$-vector model.  The other is its
supersymmetric counterpart, the so called ``hyperbolic superplane'' (see
\cite{Z90,Z91}).

The paper is organized as follows. In Sec. \ref{sec:geneq} we set up the
general description of properties of the $O(n,1)$-model in terms of a
single ``distribution function of local order parametr'' $P(\t)$. Doing
that in the spirit of dimensional continuation we consider parameter $n$ to
be any real number large enough to ensure convergence of integrals. In
Sec. \ref{sec:susy} we similarly consider supersymmetric version of our
model. Sec. \ref{sec:replim} is the central section of the paper. There
we discuss the replica limit $n \rightarrow 0$ and show that if we take it
carefully all the results for the $O(n,1)$ model {\it exactly} reproduce
results of the supersymmetric treatment. In the following sections we
proceed to solve the $O(n,1)$ model. Our analysis follows very closely
that of previous papers devoted to the problem of Anderson localization on
Bethe lattice \cite{E85-7,Z86,MF91} and therefore we omit many details.
We find that $O(n,1)$ model exhibits two phases with a phase transition
between them for any $0 \leq n < 1$. We obtain critical behaviour of
different correlators near this transition and show that it is exaclty the
same as exhibited by the supersymmetric model. Finally, Sec.
\ref{sec:dis} contains discussion of our results.

\section{$O(\lowercase{n},1)$-model: General Equations}
\label{sec:geneq}

We start with Hamiltonian
\[
{\cal H} = J \sum_{<ij>} \bn_i \cdot \bn_j  +  H \sum_{i} \s_{i}.
\]
Here $i$ and $j$ refer to the sites of Bethe lattice with coordination
number $m+1$ and $\bn=(\s,\bpi)$ is a $(n+1)$--component vector sweeping
the hyperboloid $H^{n,1}$, defined by the equation $\bn^{2} = \s^{2} -
\bpi^{2} = 1$. This hyperboloid is the symmetric space, associated with
$O(n,1)$ group: $H^{n,1} = O(n,1)/O(n)$. We parametrize $\bn$ as follows
$\s = \sqrt{1+\bpi^{2}} = \cosh \t, \quad 0 \leq \theta < \infty$,
\[ 
\bpi = \left( \begin{array}{l}
		\sinh\t \cos\phi_{1} \\ \sinh\t \sin\phi_{1} \cos\phi_{2} 
\\
\vdots \\ \sinh\t \sin\phi_{1} \sin\phi_{2} \ldots \sin\phi_{n-1}
\end{array}   \right), \quad \phi_{1}, \ldots, \phi_{n-2} \in [0, 
\pi], \quad \phi_{n-1}
\in [0, 2 \pi)  
\]
With this parametrization the scalar product is $\bn_{i}\cdot\bn_{j} = 
\s_{i} \s_{j} - \bpi_{i} \cdot \bpi_{j} \geq \cosh (\t_{i}-\t_{j}) \geq 1$
and, therefore, the hamiltonian $\cal H$ is bounded from below only for $J, 
\quad H \geq 0$. The $O(n,1)$-invariant measure on $H^{n,1}$ is
\[ 
d\bn = a \; d\t \, \sinh^{n-1}\!\t \, d\phi_{1} \sin^{n-2}\!\phi_{1} \, 
d\phi_{2} \sin^{n-3}\!\phi
_{2} \ldots d\phi_{n-1}, 
\]
where $a$ is a normalization constant to be fixed later.

Now we introduce ``distribution function of the local order parameter''
$P(\bn)$ in the usual manner.  Namely, we cut one of the $m+1$ branches
coming from site $\bn$ and integrate the part of the Boltzmann weight
$\exp(-{\cal H})$ over this branch.  The resulting function $P(\bn)$
satisfies the integral equation
\be
P(\bn) = \nint' L(\bn, \bn') D(\bn') P^{m}(\bn'),  \label{mainn} 
\ee
where we introduced the following notation:  $L(\bn, \bn') = e^{-J \bn
\cdot \bn '}, \quad D(\bn') = e^{-H \s'}$. Knowledge of the function
$P(\bn)$ allows us to calculate the partition function $Z = \nint D(\bn)
P^{m+1}(\bn)$, one-site averages $\la A(\bn) \ra = Z^{-1} \nint A(\bn)
D(\bn) P^{m+1}(\bn)$, and ``weighted'' correlators
\bea
\la A(\bn_{0})B(\bn_{r})\ra _{w} & = & \frac{N(r)}{Z} \nint_{0} A(\bn_{0}) 
D(\bn_{0})
P^{m}(\bn_{0})  \nonumber \\ 
& \times & \left( \prod_{i=1}^{r} \nint_{i} M(\bn_{i-1}, \bn_{i}) \right) 
P(\bn_{r}) B(\bn_{r}), \label{correlatorn}  \\
M(\bn, \bn') & = & L(\bn, \bn') D(\bn') P^{m-1}(\bn'),  \nonumber  
\eea
where the factor $N(r) = (m+1) m^{r-1}$ counts the number of sites located
at the distance $r$ from a given site (without this factor all the
correlators exponentially decay because of the geometry of the Bethe
lattice).

The constant $a$ in the definition of the measure $d\bn$ can be chosen
arbitrarily. It is easy to see that rescaling of the measure changes
overall normalization of $P(\bn)$ and $Z$ but does not affect either
one-site averages or correlators. This allows us to choose a convenient
normalization for $P(\bn)$ as follows. Note that when $H=0$ the Eq.
(\ref{mainn}) admits constant solution. Then we require that this 
solution be simply $P(\bn) = 1$ or, equivalently, that
\be
\nint' L(\bn, \bn') = 1.  
\label{norm}
\ee
This fixes $a = \left(\frac{J}{2 \pi}\right)^{\roo} (2 K_{\roo}(J))^{-1}$.

Magnetic field $H$ breaks the $O(n,1)$ symmetry down to $O(n)$. Then the
function $P$ may depend only on $\s$ or, equivalently, on $\t$. This 
allows us to perform angular integrations in Eq. (\ref{mainn}), yielding
\bea
P(\t) & = &\tint' L_{L}(\t, \t') D(\t') P^{m}(\t')  \label{maint},  \\
L_{L}(\t, \t') & = &\a \left( \frac{\sinh\t'}{\sinh\t} \right)^{\roo} e^{- 
\argc} \nonumber \\
& \times & (2 \pi\args)^{\frac{1}{2}} I_{\rooo}(\args),  \nonumber \\
D(\t') & = & e^{- H \cosh\t'}.  \label{draint}
\eea
Similar integration may be done in expressions for partition function $Z = 
\as \tint \sinh^{n-1}\t D(\t) P^{m+1}(\t)$, where $S_{n-1} = 
2 {\pi}^{n/2}/\G(n/2)$ is the volume of the sphere $S^{n-1}$, and 
correlators. In particular, upon averaging of $\bn$ only $\s$-component 
survives giving the ``order parameter''
\[
\la \s \ra \equiv \la \cosh\t \ra = {{\as}\over{Z}} \tint 
\sinh^{n-1}\t \cosh\t 
D(\t) P^{m+1}(\t). 
\]
For invariant correlator $\la \bn_{0} \cdot \bn_{r} \ra = \la \s_{0} \s_{r}
\ra - \la \bpi_{0} \cdot \bpi_{r} \ra $ the angular integrals give
different kernels for longitudinal $G^{L}(r) \equiv \la \cosh\t_{0}
\cosh\t_{r} \ra _{w}$ and transverse $G_{ij}^{T}(r) \equiv \la \pi_{0i}
\pi_{rj} \ra _{w}$ parts:
\bea 
G^{L}(r) & = & {{\as N(r)}\over{Z}} \tint_{0} \sinh^{n-1}\t_{0} \cosh\t_{0} 
D(\t_{0}) P^{m}(\t_{0}) \nonumber  \\
& \times & \left( \prod_{i=1}^{r} \tint_{i} M_{L}(\t_{i-1}, \t_{i}) \right)  
P(\t_{r}) \cosh\t_{r}  \label{corrL} \\
G_{ij}^{T}(r) & = & \delta_{ij} \frac{\as N(r)}{nZ} \tint_{0} 
\sinh^{n}\t_{0} D(\t_{0}) P^{m}(\t_{0}) \nonumber  \\
& \times &\left( \prod_{i=1}^{r} \tint_{i} M_{T}(\t_{i-1}, \t_{i}) \right)  
P(\t_{r}) \sinh\t_{r},  \label{corrT} 
\eea
where
\bea
M_{L}(\t, \t') & = & L_{L}(\t, \t') D(\t') P^{m-1}(\t'), \quad M_{T}(\t,
\t') = L_{T}(\t, \t') D(\t') P^{m-1}(\t'), \nonumber \\
L_{T}(\t, \t') & = &\a \left( \frac{\sinh\t'}{\sinh\t} \right)^{\roo} e^{- 
\argc} \nonumber \\
& \times & (2 \pi\args)^{\frac{1}{2}} I_{\frac{n}{2}}(\args). \nonumber
\eea

Equations (\ref{correlatorn}), (\ref{corrL}) and (\ref{corrT}) may 
be written symbolically as
$
\la A(\bn_{0}) B(\bn_{r}) \ra _{w} = Z^{-1} N(r) \la 
A(\bn_{0})|\hat{M}^{r}| 
B(\bn_{r}) \ra,
$
where $\hat{M}$ represents an integral operator with one of the kernels $M,
\quad M_{L}, \quad {\rm or} \quad M_{T}$. Introducing complete set $\mid
\phi_{\L} \ra$ of eigenfunctions of $\hat{M}$: $\hat{M} \mid \phi_{\L} \ra
= \L \mid \phi_{\L} \ra $ we can rewrite correlators as
\be
\la A(\bn_{0}) B(\bn_{r}) \ra _{w} = \frac{m+1}{m Z} \sum_{\L} (m\L)^{r} 
{{\la A \mid \phi_{\L} \ra \la \phi_{\L} \mid B 
\ra}\over{\la \phi_{\L} \mid \phi_{\L} \ra}}. 
\label{correlatorf}
\ee
We chose operators $\hat{M}$ to be non-symmetric, which means that left and 
right eigenfunction are different.

\section{Supersymmetric version: hyperbolic superplane}
\label{sec:susy}

In this section we consider the supersymmetric version of $O(n,1)$ model, 
namely, a non-linear model with field taking values on the so called 
hyperbolic superplane. This object is constructed as follows 
(see \cite{Z90,Z91}). We consider a set of 5-component vectors 
\[
{\bar{\psi}} = (\s, \pi_1, \pi_2, {\bar{\xi}}, -\xi)
\]
where the first three components are commuting, whereas the last two are
Grassmannians (we use the adjoint of the second kind, see \cite{B87} for a
review of supermathematics). Next we consider the group $G$ of linear
transformations in the space of vectors $\psi$ which preserve the
``length'' $\parallel \!\!  \psi \!\!  \parallel ^{2} = \s^{2} -
\pi_{1}^{2} - \pi_{2}^{2} - 2\bar{\xi} \xi$.  Let $K$ be the subgroup of
$G$ which separately preserves $\s^{2}$ and $\pi_{1}^{2} + \pi_{2}^{2} +
2\bar{\xi} \xi$. Then the coset space $G/K$ is isomorphic to the space of
vectors $\psi$ of unit length $\parallel \!\! \psi \!\! \parallel = 1$.
This is the hyperbolic superplane.

We will use the following parametrization of $G/K$:
\[
\pi_{1} = \sinh\t \cos\phi, \quad \pi_{2} = \sinh\t \sin\phi, \quad \s = 
\sqrt{1 + \sinh^{2}\t + 2\bar{\xi} \xi} = 
\cosh\t + \frac{\bar{\xi} \xi}{\cosh\t}.
\]
In this parametrization the $G$-invariant measure on $G/K$ is 
\[
d\psi = a \frac{1}{\s} d\pi_{1} d\pi_{2} d\bar{\xi} d\xi = a \left( 1 - 
\frac{\bar{\xi} \xi}{\cosh^{2}\t} \right) d\t \sinh\t d\phi d\bar{\xi} 
d\xi.
\]
The Hamiltonian in this case is taken to be
\[
{\cal H} = J \sum_{<ij>} \bar{\psi_{i}} \psi_{j} + H \sum_{i} \s_{i}.
\]
We again choose the constant $a$ in the definition of $d\psi$ such that
$\int \!  d\psi' \exp(-J \bar{\psi} \psi') = 1$. This gives $a =
e^{J}/2\pi$. Proceding like in section \ref{sec:geneq} we introduce
function $P(\psi)$ (by symmetry it actually depends only on $\s$) which
satisfies the equation
\[
P(\s) = \int \! d\psi' e^{-J \bar{\psi} \psi'} e^{-H \s'} P^{m}(\s').
\]
Expanding both the left-hand side and the right-hand side in powers of
Grassmann variables and integrating them out we get from the last equation
\bea
P(\t) & = & e^{J(1-\cosh\t)}e^{-H}P^{m}(0) + \tint' L_{L0}(\t, \t') D(\t') 
P^{m}(\t'),
\label{maintgr} \\
L_{L0}(\t, \t') & = & e^{J} J \sinh\t e^{-\argc} I_{1}(\args)  \label{Lo}
\eea
The first term in (\ref{maintgr}) is the boundary term resulting from 
integration by parts.

If we put $\t=0$ in Eq.  (\ref{maintgr}) we obtain $P(0) = e^{-H} P^{m}(0)$
which means that $P(0) = e^{\frac{H}{m-1}}$ or $P(0) = 0$.  To have $P(\t) 
= 1$ as a solution for $H=0$ we have to choose $P(0)=e^{\frac{H}{m-1}}$.

We can also perform Grassmann integrations in fromulae for partition
function, one-site averages and ``longitudinal'' correlators:  $Z = e^J
e^{\frac{2 H}{m-1}}, \quad \la A(\s) \ra = A(0), \quad \la A(\s_{0})
B(\s_{r}) \ra _{w} = N(r) A(0) B(0)$.  In particular we have
\be
\la \cosh\t \ra = 1, \quad G^{L}(r) \equiv \la \s_{0} \s_{r} \ra _{w} = 
N(r), 
\quad G^{L}_{c}(r) = 0,
\ee
where subscript $c$ refers to connected correlator. For transverse 
correlator $G^{T}_{ij}(r) \equiv \la \pi_{0i} \pi_{rj} \ra _{w}$ we obtain 
after some calculation
\bea
G^{T}_{ij}(r) & = & \delta_{ij} \frac{e^{J} N(r)}{Z} \tint_{0} D(\t_{0}) 
P^{m}(\t_{0}) \nonumber  \\
& \times & \left( \prod_{i=1}^{r} \tint_{i} L_{T0}(\t_{i-1}, \t_{i}) 
D(\t_{i}) P^{m-1}(\t_{i}) \right) P(\t_{r}) \sinh\t_{r}, \\
L_{T0}(\t, \t') & = & e^{J} J \sinh\t e^{-\argc} I_{0}(\args).  \nonumber
\eea

\section{Replica limit}
\label{sec:replim}

Now we study what happens with the equations of Sec. \ref{sec:geneq} for 
the $O(n,1)$-model in the replica limit when $n \rightarrow 0$, 
namely whether they reduce to those of Sec. \ref{sec:susy} or not. 

First of all, if we simply set $n=0$ in Eq. (\ref{maint}) we get $P(\t) =
\int d\t' L_{L0}(\t, \t') D(\t') P^{m+1}(\t')$ with kernel (\ref{Lo}),
which differs from Eq. (\ref{maintgr}) by the absence of the boundary
term.  From that we could conclude, in particular, that $P(0) = 0$ in the
replica limit and, therefore, this limit gives the incorrect answers.
However, this simple recipe is wrong. To see that, let us set $\t = 0$ in
Eq. (\ref{maint}) before taking the replica limit. Using small-$z$
expansion $I_{\nu}(z) \approx \frac{1}{\G(\nu + 1)} \left( \frac{z}{2}
\right) ^{\nu} $ valid for $\nu \neq -1, -2, ...$, we get
\be
P(0) = \as \tint \sinh^{n-1}\t e^{-J\cosh\t} D(\t) P^{m}(\t).
\label{P0}
\ee
If we assume that $P(0) \neq 0$ then integral in the last equation is
divergent at the lower limit if we set $n=0$. At the same time $\as = 0$
for $n=0$, so the expression (\ref{P0}) is of the type $0\cdot\infty$ in
the replica limit and should be studied in a more careful way. For this
purpose, let us consider the following identity:
\[ 
S_{n-1} \tint \sinh^{n-1}\t f(\t) = f(0) S_{n-1} \tint \sinh^{n-1}\t + 
S_{n-1} \tint \sinh^{n-1}\t (f(\t) - f(0)).
\] 
Here the first term is finite for $0<n<1$ and gives $\pi^{\roo}
\G\left(\frac{1-n}{2}\right) f(0)$, whereas in the second term the
difference $f(\t) - f(0)$ makes the integral convergent even for $n=0$.
Now we can safely take the replica limit in which the second term
disappears, and we get
\be
\lim_{n \rightarrow 0} S_{n-1} \tint \sinh^{n-1}\t f(\t) = f(0).
\label{lim1}
\ee
Therefore, the replica limit of Eq.  (\ref{P0}) is simply $P(0)=e^{-H}
P^{m}(0)$ which is exaclty the result for $P(0)$ from Sec.  \ref{sec:susy}.

Now we can perform similar trick for arbitrary $\t$:
\[ 
\tint' L_{L}(\t, \t') f(\t') = f(0) \tint' L_{L}(\t, \t') +
\tint' L_{L}(\t, \t') ( f(\t') - f(0) ).
\]
Before we take the replica limit, the integral in the first term here
equals 1 due to normalization of kernel $L(\bn, \bn')$, Eq. (\ref{norm}).
In the second term we can safely take replica limit as before. The kernel
there becomes $L_{L0}(\t, \t')$ of Eq. (\ref{Lo}). After that we split
the second term again into two pieces to get
\[ 
\lim_{n \rightarrow 0} \tint' L_{L}(\t, \t') f(\t') =
f(0) \left(1 - \tint' L_{L0}(\t, \t')\right) + \tint' L_{L0}(\t, \t') 
f(\t').
\]
The integral in the first term can be done (see, e.g. \cite{PBM}) and we 
get as the result
\be  
\lim_{n \rightarrow 0} \tint' L_{L}(\t, \t') f(\t') = 
f(0) e^{J(1-\cosh\t)} + \tint' L_{L0}(\t, \t') f(\t').
\label{lim2}
\ee
Then in the replica limit Eq. (\ref{maint}) becomes exactly the 
Eq. (\ref{maintgr}) of Sec. \ref{sec:susy}!

Alternative way of getting Eq. (\ref{lim2}) is to separate the two
contributions to the modified Bessel function which enters the integral
kernel $L_L(\theta,\theta')$ using the recursion relation
\[
I_{\frac{n}{2} -1}(z) = \frac{n}{z} I_{\frac{n}{2}}(z) + I_{\frac{n}{2}+1} 
(z) 
\simeq {1\over \Gamma(n/2)}\left({z\over 2}\right)^{{n\over 2}-1} + I_1(z)\ 
;\quad n\to 0
\]
The crucial point is that the first term here cannot be neglected,
although it has a vanishing coefficient in the limit $n\to 0$, since
the corresponding integral over $\theta'$ will diverge in this limit. 
Thus, in full analogy with the treatment of Eq. (\ref{P0}), this
singular contribution should be first evaluated at finite $n$, and
only then the limit $n\to 0$ can be taken. This again gives the 
Eq. (\ref{lim2}).

The use of prescriptions (\ref{lim1}) and (\ref{lim2}) allows us to show
that equations for all the quantities of interest from the Sec.
\ref{sec:geneq} in the replica limit reduce to those of the supersymmetry
method of Sec. \ref{sec:susy}.  This is the main result of our analysis.
In the remaining sections we find that our model exhibits ordered and
disordered phases for $0 \leq n <1$.  We find the transition point between
them and solve for the critical behaviour of correlators near this
transition.  We explicitly show then that in the replica limit this
behaviour is identical to the one found in Refs. \cite{E85-7,Z86} for 
supersymmetric $\sigma$-model and in \cite{MF91} for the Anderson model 
on Bethe lattice.

\section{Ordered phase and transition point}

For $J \gg 1$ we expect an ordered state with spontaneuosly broken
symmetry, where all the $\bn$'s are slightly fluctuating around
$\s$-direction. The transverse components $\bpi$ are small and we can
expand in them: $\s \approx 1 + \frac{1}{2} \bpi^2 + \cdots$.  Then the
kernel of the Eq. (\ref{mainn}) becomes Gaussian and, therefore, the
equation admits (for any value of $n$) a solution which is also Gaussian:
\be
P(\bn) = c e^{-\frac{1}{2} \alpha \bpi^2},
\label{gaussianp}
\ee
where $\alpha = \frac{1}{2 m} \left( J(m-1) - H + \sqrt{(J(m-1) - H)^2 + 4 
m J H} \right)$ $(\alpha = J \frac{m-1}{m}\ \mbox{for}\ H = 0)$ 
and $c$ is some normalization constant. With this form of $P(\bn)$ we 
perform Gaussian integrals to find for small magnetic field
\be
G^{L}_{c}(r) \propto n m^{-r}, \quad G^{T}_{ij}(r) = 
\delta_{ij} \frac{1}{2(m-1)\rho_{s}} 
\exp \left(- \frac{H}{(m-1)\rho_{s}} r \right), 
\label{corrsorder}
\ee
where ``spin stiffness'' $\rho_{s} = J$. So longitudinal correlator is
massive, while transverse modes are Goldstones with long-ranged 
correlations for $H = 0$.

Non-compact nature of the $O(n,1)$ model makes it rather unusual from the
traditional point of view. For example, mean field theory for this model
does not give any transition at all: the system is always ordered. If we
expect to have disordered phase for $J \ll 1$, then in this phase the order
parameter will be large for small magnetic field, diverging when $H
\rightarrow 0$.

The symmetry breaking factor $D(\t)$, Eq. (\ref{draint}), becomes
significant for $\cosh\t \sim H^{-1}$ or $\t \sim \ln \frac{2}{H}$,
therefore it is convenient to change variables to $t = \ln (H \cosh\t)$ in
Eq. (\ref{maint}). After such a change the argument of Bessel function
becomes large for small $H$, and, using asymptotic formula $(2\pi z)^{1/2}
I_{\nu}(z) \rightarrow e^{z}$ we arrive in the limit $H \rightarrow 0$ at
\bea
P(t) & = & \intt' L(t-t')D(t')P^{m}(t'),  \label{mainasympt} \\
L(t) & = & \a \exp\left(\frac{1-n}{2} t -J \cosh t \right), \quad D(t) = 
\exp(-e^{t}). 
\nonumber
\eea
This equation was derived under the assumption that the solution $P$ is a
function of variable $H\cosh\t$ for small $H$.  Such a solution becomes
$P=1$ in the limit $H=0$.  Solution of the ``ordered 
type'' (\ref{gaussianp}) is not contained in the Eq. (\ref{mainasympt}) or,
rather, it corresponds to trivial solution $P=0$.

The qualitative behaviour of the solution of the Eq. (\ref{mainasympt})
should be as follows. For large positive $t$ ($t \gg 0$) the function
$P(t)$ exponentially goes to zero because of the symmetry breaking term
$D(t')$. For large negative $t$ ($t \ll 0$) the main contribution to the
integral in Eq. (\ref{mainasympt}) comes from $t' \ll 0$ because the
kernel $L(t-t')$ is sharply peaked at $t=t'$. But in this region $D(t')
\approx 1$ and the Eq. (\ref{mainasympt}) admits solution $P=1$. In the
region $t \sim 0$ there should be a kink connecting the two asymptotic
solutions. The precise form of the solution can be found numerically by
iterations starting with $P=1$. But such iterative procedure is convergent
to the solution of described type only for small enough $J$. If $J >
J_{c}$, where $J_{c}$ is the critical coupling, this solution becomes
unstable, the kink at $t \sim 0$ starts to move to the negative $t$ until
the solution $P(\t)$ takes the form characteristic of the ordered phase.
In fact, Eq. (\ref{mainasympt}) is very similar to the one studied by
Zirnbauer \cite{Z86}, so we follow very closely the stability analysis of
this paper.

For $t \ll 0$ we drop $D(t')$ and linearize the simplified equation around 
constant solution: $P(t) = 1 - \delta P(t)$ with
\be
\delta P(t) = m \intt' L(t-t') \delta P(t')
\label{mainlint}
\ee
Because of the translational invariance of this equation it admits 
exponential solutions: $\intt' L(t-t') e^{\nu t'} = \L(\nu) e^{\nu t}$, 
where 
\be 
\L(\nu) = \frac{K_{\roo + \nu}(J)}{K_{\roo}(J)}
\label{eigenvalue}
\ee
is an eigenvalue of the integral operator with the kernel $L(t-t')$.

Let us make a mathematical remark. In terms of vector $\bn$ 
Eq. (\ref{mainlint}) corresponds to $\delta P(\bn) = m \nint' L(\bn, \bn')
\delta P(\bn')$, i.e., $\delta P(\bn)$ is an eigenfunction of an
$O(n,1)$-invariant integral operator $\hat{L}$ with kernel $L(\bn, \bn')$.
The space of such functions is spanned by the so called spherical functions
of group $O(n,1)$ (see, e. g. \cite{VK93}). In our case of
$O(n)$-symmetric $P(\t)$ these are the zonal spherical functions
\be
\psi_{\nu}(\t) = (\sinh\t)^{1-n/2} P_{\nu + n/2 -1}^{-n/2 +1}(\cosh\t),
\label{sphericalfns}
\ee
where $P_{\nu}^{\mu}(z)$ is the Legendre function.  The funciton
$\psi_{\nu}(\t)$ is an eigenfunction of $\hat{L}$ with an eigenvalue
$\L(\nu)$ given by Eq. (\ref{eigenvalue}). Some properties of these
functions are presented in the Appendix.

For the problem of stability of the asymptotic solution $P=1$ the relevant
values of $\nu$ are real positive numbers. Indeed, the natural
perturbation $\delta P(t)$ induced by the symmetry breaking term $D(t)
\approx 1 - e^{t}$ in the region $t \ll 0$ is $\delta P(t) = e^{\nu t}$
with $\nu = 1$. Analysis similar to that of \cite{Z86} shows that if $m 
\L_{min} (\nu) > 1$ then the solution $P=1$ is unstable and
collapses to the trivial solution $P=0$ (ordered phase). This happens for
any value of coupling constant $J$ for $n \ge 1$ because in this case
$K_{\roo + \nu}(J) \ge K_{\roo}(J)$ and $m \L_{min}(\nu) = m
\L(0) = m > 1$. On the other hand, for $0 \le n < 1$, $K_{\roo}(J) =
K_{|\roo|}(J)$, and $\ro +\nu$ may be smaller than $|\ro|$. In this case
there is a transition at the value of the coupling constant $J_{c}$
determined by the equation
\be 
m \frac{K_{0}(J_{c})}{K_{|\roo|}(J_{c})} = 1.
\label{transitionpoint}
\ee
Solution of this equation for $m = 2$ is shown on Fig. 1.

\section{Correlators in the disordered phase}

From now on we restrict our attention to the case $0 \le n <1$. We will
also assume that magnetic field is very small: $H \ll 1$. In this case we
can approximate $P(0)$ by 1. Then for $J < J_{c}$ function
\[
P(\t) \approx \left\{ \begin{array}{ll}
1, & \t \ll \ln \frac{2}{H}  \\
0, &  \t \gg \ln \frac{2}{H}  \end{array} \right.
\]
with a kink connecting these two regions near $\t = \ln \frac{2}{H}$.
Typical solution of Eq. (\ref{maintgr}) of this type for $n=0$, 
$J=10^{-6}$, $H=10^{-10}$ is shown on Fig. 2. Approximating this
solution by a step function we estimate partition function as
\[
Z = \as \int_{0}^{\ln \frac{2}{H}} \! d\t \sinh^{n-1}\t,
\]
which is not singular as $H \rightarrow 0$, so we'll calculate it for 
$H=0$:
\be
Z = \as \tint \sinh^{n-1}\t = a \pi^{\roo} \G\left(\frac{1-n}{2}\right).
\label{partfn}
\ee
The finitness of this quantity reflects the fact that the total volume of 
the hyperboloid $H^{n,1}$ is finite for $0 \le n <1$. We also get the 
``order parameter''
\be
\la \s \ra \approx \frac{\as}{Z} 
\int_{0}^{\ln \frac{2}{H}} \! d\t \sinh^{n-1}\t 
\cosh\t \approx \frac{\as}{nZ} H^{-n}.
\label{sigma}
\ee
So $\la \s \ra$ diverges when $H \rightarrow 0$ in the disordered phase, 
unless $n=0$, in which case $\la \s \ra$ becomes non-critical (similar 
to the density of states in Anderson transition).

To obtain the expression for correlators we again perform the change of
variables $H \cosh\t = e^{t}$ in Eqs. (\ref{corrL}), (\ref{corrT}). To
the leading order in $1/H$ both $G^{L}$ and $G_{ij}^{T}$ become the same
(up to a factor $\delta_{ij}/n$):
\[
G^{L}(r) = \frac{\as N(r)}{Z} H^{-n-1} \intt_{0} e^{nt_{0}} D(t_{0}) 
P^{m}(t_{0}) \left( \prod_{i-1}^{r} \intt_{i} M_{L}(t_{i-1}, t_{i}) \right) 
P(t_{r}) 
e^{t_{r}}. 
\]

Range of integration for operator with kernel $M_{L}(t, t')$ is infinite,
which means that the spectrum of eigenvalues is continuous. In this case
Eq. (\ref{correlatorf}) for the correlator $G(r)$ is
\[
G^{L}(r) = \frac{(m+1) \as}{m Z} H^{-n-1} \int_{0}^{\infty} \! d\l W(\l) 
A^{2}(\l) 
(m \L_{\l})^{r},
\]
where $A^{2}(\l) = \la e^{n t} | \phi^{r}_{\l}(t) \ra \la \phi^{l}_{\l}(t)
| e^{t} \ra $ and $W(\l)$ comes from the normalization of eigenfunctions.
Right and left eigenfunctions $\phi^{r,l}_{\l}(t)$ in the limit $H
\rightarrow 0$ should approach $\psi_{\frac{1-n}{2} + i\l}(\t)$ and
$\sinh^{n-1}\t \psi_{\frac{1-n}{2} + i\l}(\t)$, or asymptotically for $t
\ll 0$
\be 
\phi^{r,l}_{\l}(t) 
\propto \frac{e^{\mp \ro t}}{\l} \sin\left( \l t + \delta(\l) 
\right)
\label{phi}
\ee
with eigenvalue $\L_{\l} = \frac{K_{i\l}(J)}{K_{\roo}(J)}$. As is
suggested by Eqs. (\ref{normalization}) and (\ref{smalllmeasure}) the
normalization of $\phi^{r,l}_{\l}(t)$ should be
\[
\intt \phi^{l}_{\l}(t) \phi^{r}_{\l'}(t) = \frac{\delta(\l - \l')}{W(\l)},
\]
where $W(\l) \propto \l^{2}$ for small $\l$. Small $\l$ behaviour of
``phase shift" $\delta(\l)$ in Eq. (\ref{phi}) is found matching the
asymptotic behaviour (\ref{phi}) to $\phi_{\l}(t) = 0$ for $t > 0$, which
is again the effect of the term $D(t)$. Such matching gives that
$\delta(\l)$ should be at least linear in $\l$ for small $\l$. Using this
we can show that $A(\l) \sim O(1)$ for small $\l$. Also expanding $m
\L_{\l}$ in $J_{c} - J$ and in $\l$ we find $m \L_{\l} \approx 1 - a_{0}
(J_{c} - J) - a_{2} \l^{2}$. Combining all the above results we arrive at
\bea
G^{L}(r) & \propto & n H^{-n-1} \int_{0}^{\infty} \! d\l \l^{2} e^{- a_{0} 
(J_{c} - J) r - a_{2} \l^{2} r} \propto n H^{-n-1} r^{-3/2} e^{-r/\xi}, 
\nonumber \\
G^{T}_{ij}(r) & \propto & \delta_{ij} H^{-n-1} r^{-3/2} e^{-r/\xi},
\nonumber
\eea
where $\xi \sim (J_{c} - J)^{-1}$.

In the replica limit these equations reduce to $G^{L}(r) = 0$ and
\[
\G^{T}_{ij}(r) \equiv \lim_{H \rightarrow 0} H G^{T}_{ij}(r) \propto
\delta_{ij} r^{-3/2} e^{-r/\xi},
\]
which has  the same form as the density--density correlator in
the localized regime in Refs.\cite{E85-7,Z86,MF91}.

\section{Correlators in the ordered phase close to the transition}

In the case $J-J_{c} \ll J_{c}$ (just above the transition) we expect
spontaneous symmetry breaking which in terms of function $P(\t)$ takes 
place at some large scale $A$ divergent at $J=J_{c}$ such that
\[
P(\t) \approx \left\{ \begin{array}{ll}
1, & \t \ll \ln 2A  \\
0, &  \t \gg \ln 2A.  \end{array} \right.
\]
Similarly to the previous section we find that partition function is 
nonsingular as $A \rightarrow \infty$, so we take it to be equal to its 
value at $J_{c}$, Eq. (\ref{partfn}). Again, like in Eq. (\ref{sigma}) we 
find
\be
\la \s \ra \approx \frac{\as}{nZ} A^{n}.
\ee

The long distance behaviour of the correlators in this phase is determined 
by the eigenvalues of the operator with kernel $M(\bn, \bn')$, see 
Eq. (\ref{correlatorn}). Because of the form of the function $P(\t)$ the 
integration range in $\t$ for this operator is finite, and the spectrum of 
its eigenvalues is discrete. It is easy to see that for $H=0$ the largest 
eigenvalue of this operator is $1/m$. Indeed, for $Q \in O(n,1)$ using 
invariance of the kernel $L(\bn, \bn')$ and the measure $d\bn$ we get 
$P(Q\bn) = \nint' L(\bn,\bn') P^{m}(Q\bn')$. Taking $Q$ to be 
infinitesimally close to unity and expanding we obtain
\be
\frac{dP(\bn)}{d\s} \bpi = m \nint' L(\bn, \bn') P^{m-1}(\bn')
\frac{dP(\bn')}{d\s'} \bpi'.
\label{goldstones}
\ee  
The functions $|f_{i} \ra = \frac{dP(\bn)}{d\s} \pi_{i}$ are the Goldstone 
modes associated with the symmetry breaking. Integrating out angular 
variables in (\ref{goldstones}) we have
\be
\frac{dP(\t)}{d\t} = m \tint' L_{T}(\t, \t') P^{m-1}(\t') 
\frac{dP(\t')}{d\t'}.
\label{eqndp}
\ee
In the asymptotic region $1 \ll \t, \t' \ll \ln 2A$ the kernel 
$L_{T}(\t, \t') \rightarrow L(\t-\t')$ of Eq. (\ref{mainasympt}), 
Eq. (\ref{eqndp}) becomes the same as Eq. (\ref{mainlint}) which means that 
$P'(\t)$ has the asymptotic behaviour 
\be
\frac{dP(\t)}{d\t} \approx C \frac{e^{\frac{1-n}{2} \t}}{\l} \sin \l \t
\label{dp}
\ee
where $\l$ satisfies $\frac{K_{i\l}(J)}{K_{\roo}(J)} = \frac{1}{m}$. 
Expanding this in small $\l$ and small $J-J_{c}$ we find 
\be
\l \sim (J-J_{c})^{1/2} .
\label{lambda}
\ee
The fast decrease of $P^{m-1}(\t)$ near $\t = \ln 2A$ chooses the value of 
$\l$ such that function (\ref{dp}) has the first node a this point: $\l = 
\frac{\pi}{\ln 2A}$. Combining this with (\ref{lambda}) we find 
\be
A \sim \exp \left( {\rm const} (J-J_{c})^{-1/2} \right).
\label{a}
\ee

The value of the constant $C$ in Eq. (\ref{dp}) may be found as follows.
$dP/d\t$ is related to the function $\delta P(\t)$ of Eq. (\ref{mainlint})
in an obvious manner: $\frac{dP}{d\t} = - \frac{d\delta P}{d\t}$. Using
this and writing $\delta P(\t) \approx C_{1} \frac{e^{\frac{1-n}{2}
\t}}{\l} \sin (\l \t + \delta (\l))$ we get $\delta(\l) = \tan^{-1} \left(
\frac{2\l}{n-1} \right) \approx \frac{2\l}{n-1}$ and $C = - C_{1}
\sqrt{\l^{2} + \left(\ro\right)^{2}} \approx - C_{1} |\ro|$. But function
$\delta P$ should approach the value 1 near $\t = \ln 2A$. That gives
$C_{1} \approx |\ro| (2A)^{\roo}$ and
\be
C \approx - \left(\ro\right)^{2} (2A)^{\roo}.
\label{C}
\ee

Now we evaluate the largest eigenvalue $\L_{max}$ of operator with kernel
$M(\bn, \bn')$ for small $H$ using the first order perturbation theory:
$\L_{max} = { \la f_{i} | {\hat{M}} | f_{i} \ra } /{\la f_{i} | f_{i} \ra
}$ (no summation!). Since kernel $M(\bn, \bn')$ is non-symmetric, $\la
f_{i}|$ differs from $ |f_{i} \ra$ by the factor $P^{m-1}(\bn)$. Then we
have
\[
\la f_{1} | f_{1} \ra = \nint P^{m-1}(\bn) \left( \frac{dP}{d\s} 
\pi_{1} \right) ^{2} = \frac{\as}{n} \tint \sinh^{n-1} \t P^{m-1}(\t) 
\left( \frac{dP(\t)}{d\t}  \right)^{2}.  
\]
Using Eqs. (\ref{dp}) and (\ref{C}) we estimate $\la f_{1} | f_{1} \ra$ to 
be 
\[
\la f_{1} | f_{1} \ra \propto \frac{\as}{n}
\frac{C^{2}}{\l^{2}} \int_{0}^{\ln 2A} \! d\t \sin^{2} \l\t 
\propto \frac{\as}{n} A^{n-1} \ln^{3} 2A.
\]
In the presence of $H$ Eq. (\ref{goldstones}) is replaced by
\[
\frac{dP(\bn)}{d\s} \bpi = \nint' L(\bn, \bn') 
\frac{d}{d\s'} \left( P^{m}(\bn') 
e^{-H\s'} \right) \bpi'.
\]
Using this we perform integrations by parts and keep only linear terms in 
$H$ to find
\[
\la f_{1} | {\hat{M}} | f_{1} \ra = \frac{\la f_{1} | f_{1} 
\ra }{m} - H \frac{Z \la \s \ra}{m^{2}(m+1)}.  
\]
Then for the maximal eigenvalue we have $ m \L_{max} = 1 
- H/((m-1)\rho_{s})$, where the ``spin stiffness'' 
\be
\rho_{s} = \frac{m(m+1)}{(m-1)} \frac{\la f_{1} | f_{1} \ra}{Z \la \s \ra} 
\propto \frac{\ln^{3}2A}{A} \sim 
(J-J_{c})^{-3/2} \exp \left( - {\rm const} (J-J_{c})^{-1/2} \right).
\label{spinstiff}
\ee

For the final evaluation of the correlator $\la A(\bn_{0}) B(\bn_{r}) \ra
_{w}$ we have to calculate $A^{2}(\L) = \sum_{i=1}^{n} \la A | f_{i} \ra
\la f_{i} | B \ra / \la f_{1} | f_{1} \ra$. For longitudinal correlator
$A(\bn) = B(\bn) = \cosh\t$ in which case $\la f_{i} | \cosh\t \ra = \nint
P^{m}(\bn) \s \frac{dP(\bn)}{d\s} \pi_{i} = 0$ and $A^{2}(\L) =0$. This
means that the decay of $G^{L}(r)$ is governed by smaller eigenvalues and,
therefore, $G^{L}(r)$ is massive even for $H=0$, which is to be compared
with Eq. (\ref{corrsorder}). On the other hand, for Goldstone modes we
have $\la f_{i} | \pi_{j} \ra = \nint P^{m}(\bn) \frac{dP(\bn)}{d\s}
\pi_{i} \pi_{j} = - \delta_{ij} \frac{Z \la \s \ra}{m+1}$ and for the
transverse correlator we finally have
\be
G_{ij}^{T}(r) = \delta_{ij} \frac{m+1}{mZ} A^{2}(\L) (m\L_{max})^{r} = 
\frac{\la \s \ra}{(m-1)\rho_{s}} \exp 
\left( -\frac{H}{(m-1)\rho_{s}} r \right).
\label{tcorr}
\ee

Note that again, as before, in the replica limit our equations (\ref{a}), 
(\ref{spinstiff}) and (\ref{tcorr}) exactly correspond to the results of 
Refs. \cite{E85-7,Z86,MF91}.

\section{Discussion}
\label{sec:dis}

In this paper, we have solved the non-compact $O(n,1)$ model on the Bethe
lattice for arbitrary $n$. The analytical continuation procedure allows us
to consider $n$ as being an arbitrary positive number. We find that for
$n>1$ the symmetry of the model is always broken, so that the system is
ordered and the order parameter has a finite value. This is in agreement
with Ref. \cite{CR83}, where the system was shown to be ordered for $n=1$
and $n\to\infty$ above two dimensions. (Note that the Bethe lattice is
effectively infinite dimensional). However, for $0<n<1$ we find a
transition from the ordered to disordered phase, when the coupling strength
$J$ decreases. In the replica limit $n\to 0$, our solution reproduces
exactly that for the supersymmetric version of the model. The latter is
nothing but the hyperbolic superplane introduced by Zirnbauer \cite{Z90,Z91} 
as a toy model for the Anderson localization transition. In the whole
region $0\le n<1$ the qualitative picture of the transition and the
critical behavior are analogous to those of the supersymmetric model of
Zirnbauer, which shows in turn all the essential features of the Anderson
transition on Bethe lattice studied in Refs. \cite{E85-7,Z86,MF91}. The
success of the replica trick may seem surprising, since we know \cite{VZ85}
that it gives wrong result in the case of the level correlation problem.
The crucial difference is that near the Anderson transition only the
non-compact sector of the supersymmetric $\sigma$--models is essential,
whereas for the level correlation problem both compact and non-compact
sectors are equally important. The similar reason explains the agreement
between the replica trick renormalization group calculation of asymptotics
of various distributions \cite{AKL} and the recent study of these
asymptotics via the supersymmetry method \cite{M95}.

When the manuscript was in preparation, we learnt about the work by
T. Dupr\'e \cite{D95} who studied the supersymmetric model of Zirnbauer
\cite{Z90,Z91} in 3D and found the critical behavior analogous to that
expected for the Anderson localization transition.

\section*{Acknowledgements}

We are grateful for the hospitality extended to us during the Spring
College in Condensed Matter in May 1994 in the International Center for
Theoretical Physics (Trieste), where the main part of this work was done.
One of us (I.A.G.) deeply appreciates discussions with N. Read, who
suggested studying non-compact models in the first place, and R. Shankar.
The financial support of the Alexander von Humboldt Foundation and SFB 195
der Deutschen Forschungsgemeinschaft (A.D.M.) and of NSF grant 
No. DMR-91-57484 (I.A.G.) is acknowledged with thanks.

\appendix
\section*{}

The functions (\ref{sphericalfns}) have the following asymptotic behavior
\be
\psi_{\nu}(\t \gg 1) \approx \frac{2^{n/2 -1}}{\sqrt{\pi}} \left( 
\frac{\G(\nu + \ro)}{\G(\nu + n - 1)} e^{\nu \t} + \frac{\G(-\nu - 
\ro)}{\G(-\nu)} e^{-(\nu + n -1) \t} \right).
\label{largeasymp}
\ee

Among the functions (\ref{sphericalfns}) there is a special subset with
$\nu = \frac{1-n}{2} + i\l$ , $\l \ge 0$. These functions are real and
form a continuous basis in the space $L^{2}\left( [1, \infty), d\t
\sinh^{n-1}\t \right)$:
\be
\tint \sinh^{n-1}\t \psi_{\frac{1-n}{2} + i\l}(\t) \psi_{\frac{1-n}{2} + 
i\l'}(\t) = 
\frac{1}{\mu(\l)} \delta (\l-\l'),
\label{normalization}
\ee
\be
\lint \psi_{\frac{1-n}{2} + i\l}(\t) \psi_{\frac{1-n}{2} + i\l}(\t') = 
\sinh^{1-n}\t \delta(\t - 
\t'),
\label{completeness}
\ee
\be
\mu(\l) = \left| \frac{\G(\ro + i\l)}{\G(i\l)} 
\right| ^{2}.
\label{spectralmeasure}
\ee
For $n\ne 1$ and for small values of $\lambda$, the 
asymptotics (\ref{largeasymp}) and the spectral measure 
(\ref{spectralmeasure}) take the form
\be
\psi_{\frac{1-n}{2} + i\l}(\t) \approx \frac{2^{n/2}}{\sqrt{\pi} 
\G\left(\ro\right)} 
\frac{e^{\frac{1-n}{2} \t}}{\l} \sin \l\t, \quad
\mu(\l) \approx \G^{2}\left(\ro\right) \; \l^{2}. 
\label{smalllmeasure}
\ee

\begin{figure}[t]
\epsfbox[0 100 10 370]{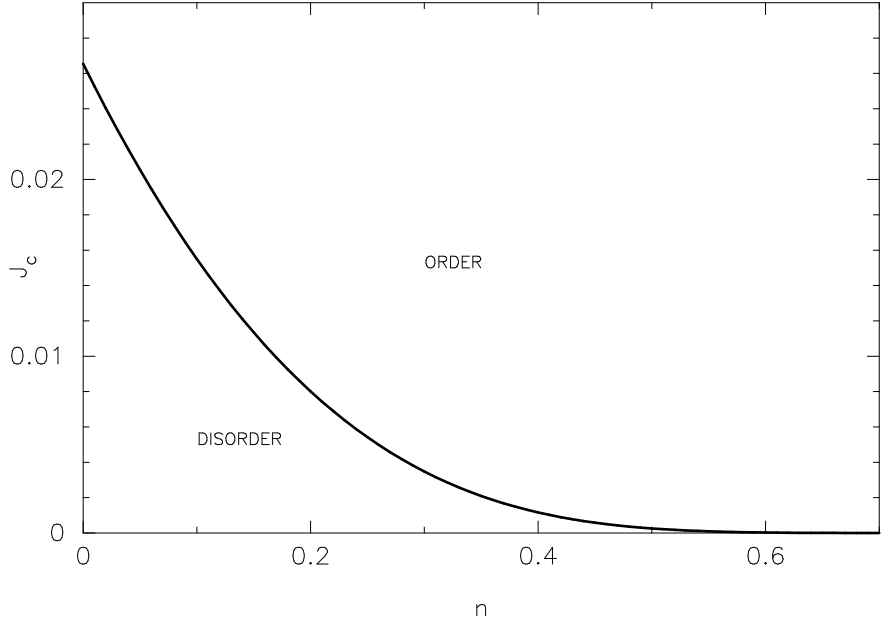}
\caption{Solution of 
Eq. (\protect{\ref{transitionpoint}}) (phase diagram on 
the $J - n$ plane) for $m = 2$.}
\label{fig1}
\end{figure} 

\begin{figure}[b]
\epsfbox[0 100 10 370]{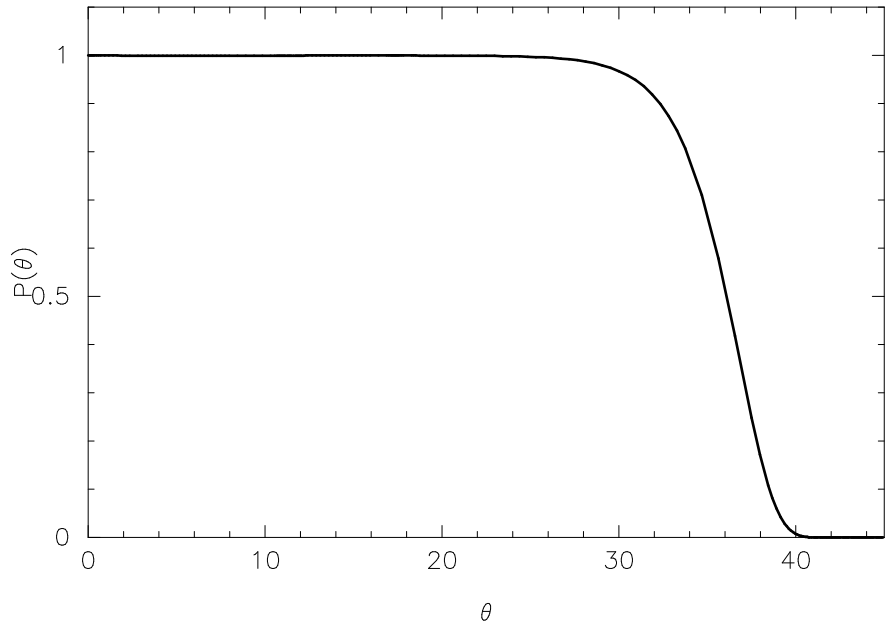}
\caption{Function $P(\t)$ for $n = 0, \quad J= 10^{-6}, \quad H=10^{-10}$ 
obtained by
numerical 
{\protect \linebreak}
solution of Eq. (\protect{\ref{maintgr}}).}
\label{fig2}
\end{figure}

\end{document}